\documentclass[fleqn,usenatbib]{mnras}

\usepackage{newtxtext,newtxmath}

\usepackage[T1]{fontenc}
\usepackage{ae,aecompl}

\usepackage{graphicx}	

\newcommand{\kepler}{{\em Kepler\/}}
\newcommand{\tess}{{\em TESS\/}}
\newcommand{\dscuti}{\texorpdfstring{\mbox{$\delta$~Scuti}}{delta Scuti}}

\newcommand{\cd}{\mbox{d$^{-1}$}}

\newcommand{\echelle}{{\'e}chelle}

\newcommand{\bprp}{\mbox{$G_{BP}-G_{RP}$}}
\newcommand{\ngaia}{2844}
\newcommand{\ntess}{1708}
\newcommand{\ndsct}{848}

\newcommand{\nbarac}{304}
\usepackage{academicons}
\usepackage{svg}
\newcommand{\orcid}[1]{\href{https://orcid.org/#1}{\textsuperscript{\includegraphics[width=10pt]{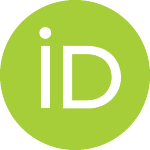}}}}
\usepackage{CJKutf8}
\newcommand{\CNnames}[1]{{\begin{CJK}{UTF8}{gbsn}~(#1)~\end{CJK}}}

\usepackage{longtable}
\usepackage{booktabs} 
\graphicspath{{./}{figures/}}

\newif\ifarxiv
\arxivtrue 

\newcommand{\new}[1]{{\color{red}\textbf{#1}}}
\renewcommand{\new}[1]{\relax{#1}}
\newcommand{\newtwo}[1]{{\color{red}\textbf{#1}}}
\renewcommand{\newtwo}[1]{\relax{#1}}

\title[Identifying 850 $\delta$ Scuti stars]{Identifying 850 $\delta$ Scuti pulsators in a narrow Gaia colour range with \tess\ 10-minute full-frame images}

\author[Read et al.]{%
Amelie K. Read$^1$\thanks{E-mail: area9433@uni.sydney.edu.au}, 
Timothy R. Bedding\orcid{0000-0001-5222-4661},$^1$\thanks{E-mail: tim.bedding@sydney.edu.au}
Prasad Mani\orcid{0000-0002-8707-201X},$^1$
Benjamin T. Montet\orcid{0000-0001-7516-8308},$^{2,3}$
\newauthor
Courtney Crawford\orcid{0000-0002-7654-7438},$^1$
Daniel R. Hey\orcid{0000-0003-3244-5357},$^4$
Yaguang~Li\CNnames{李亚光}\orcid{0000-0003-3020-4437},$^{1,4}$
\newauthor
Simon J. Murphy\orcid{0000-0002-5648-3107},$^{5}$
May Gade Pedersen\orcid{0000-0002-7950-0061}$^1$ and
\new{Joachim Kruger}\orcid{0009-0003-3841-5383}$^{5}$
\newauthor
\\
$^1$Sydney Institute for Astronomy, School of Physics, University of Sydney, NSW 2006, Sydney, Australia \\
$^2$School of Physics, University of New South Wales, NSW 2052, Australia\\
$^3$UNSW Data Science Hub, University of New South Wales, Sydney, NSW 2052, Australia\\
$^4$Institute for Astronomy, University of Hawai`i, Honolulu, HI 96822, USA\\
$^5$Centre for Astrophysics, University of Southern Queensland, Toowoomba, QLD 4350, Australia\\
}

\date{}

\pubyear{2023}

\begin{document}
\label{firstpage}
\pagerange{\pageref{firstpage}--\pageref{lastpage}}
\maketitle

\begin{abstract}
We use \tess\ 10-minute Full Frame Images (Sectors 27--55) to study a sample of \ntess\ stars within 500\,pc of the Sun that lie in a narrow colour range in the centre of the \dscuti\ instability strip ($0.29 < \bprp < 0.31$).  Based on the Fourier amplitude spectra, we identify \ndsct\ \dscuti\ stars, as well as 47 eclipsing or contact binaries.
The strongest pulsation modes of some \dscuti\ stars fall on the period--luminosity relation of the fundamental radial mode but many correspond to overtones that are approximately a factor of two higher in frequency. Many of the low-luminosity \dscuti\ stars show a series of high-frequency modes with very regular spacings. 
The fraction of stars in our sample that show \dscuti\ pulsations is about 70\% for the brightest stars ($G<8$), consistent with results from \kepler. However, the fraction drops to about 45\% for fainter stars and we find that a single sector of \tess\ data only detects the lowest-amplitude \dscuti\ pulsations (around 50\,ppm) in stars down to about $G=9$.
\newtwo{Finally, we have found four new high-frequency \dscuti\ stars with very regular mode patterns, and have detected pulsations in $\lambda$~Mus that make it the fourth-brightest \dscuti\ in the sky ($G=3.63$).}
Overall, these results confirm the power of \tess\ and Gaia for studying pulsating stars.
\end{abstract}





\section{Introduction}

The Transiting Exoplanet Survey Satellite (\tess) is providing all-sky time-series space photometry at high cadence. In addition to pre-selected targets that may be observed at 120-s or 20-s cadence, the spacecraft takes Full Frame Images (FFIs). The cadence of the FFIs has been decreasing throughout the mission: they began at 30-min cadence, reduced to 10-min cadence, and are now being taken every 200\,s. With these FFIs one can create custom light curves for any star falling on a TESS camera during any TESS observing sector.

One class of pulsating stars that has benefited enormously from the plentiful \tess\ data are the $\delta$~Scutis. These A/F type stars are abundant and are bright enough for high-quality TESS photometry. They pulsate with periods between $\sim$20\,min and 5\,hr, hence even the highest-frequency pulsators are well sampled by the 10-minute FFIs. Further, with coherent oscillations that are only weakly damped, a single sector is sufficient to fully characterise them. Additional sectors are helpful to study any secular behaviour and to improve the overall signal-to-noise to detect any weak modes.
Results on \dscuti\ stars using \tess\ have included studies of individual stars too numerous to list,
as well as several studies of larger ensembles of pulsators 
\citep[e.g.][]{Antoci++2019, Balona++2019, Balona+Ozuyar2020, Barac++2022, Skarka++2022, Bedding2023, Daszyska-Daszkiewicz++2023, Li-Gang++2023, Palakkatharappil+Creevey2023, Pamos-Ortega++2023, Xue++2023}.

This paper looks at \tess\ photometry of stars in a narrow vertical strip in the colour-magnitude diagram that lies in the centre of the \dscuti\ instability strip.  This is a pilot study, which we plan to extend to a full all-sky survey. We make use of the 10-min FFIs that were taken in years 3 and 4 of the mission, which have a higher Nyquist frequency than the 30-min FFIs and have not yet been exploited for \dscuti\ analyses.

\section{Methods and Results}

\subsection{Sample selection}
\label{sec:sample}

We selected stars from the Gaia DR3 catalogue \citep{Gaia++2021} in a narrow colour range: $0.29 < \bprp < 0.31$. 
This colour selection lies \new{approximately} in the centre of the \new{observed} \dscuti\ instability strip \new{\citep{Bedding2023}} and thus we expect a significant proportion of stars to be pulsators. We limited our sample to stars within 500\,pc of the Sun (parallax $>$2\,mas) and apparent magnitude $G< 12$ (to exclude white dwarfs).  Absolute magnitudes were calculated from Gaia DR3 apparent magnitudes and parallaxes, and no corrections were made for extinction or reddening.

To exclude Gaia solutions with spurious parallaxes, we removed 65 stars whose astrometric reliability diagnostic, \texttt{fidelity\_v2}, calculated by \citet{Rybizki++2022} was less than 0.75 (which was the threshold adopted by \citealt{El-Badry++2022}).
We also excluded two stars with spurious Gaia photometry (HD~79613B and HD~161740B), based on their values of \texttt{phot\_bp\_rp\_excess\_factor} \citep{Riello++2021} being much greater than 1 (the values are 70.2 and 8.5, respectively).







\begin{figure}
\includegraphics[width=1.0\linewidth]{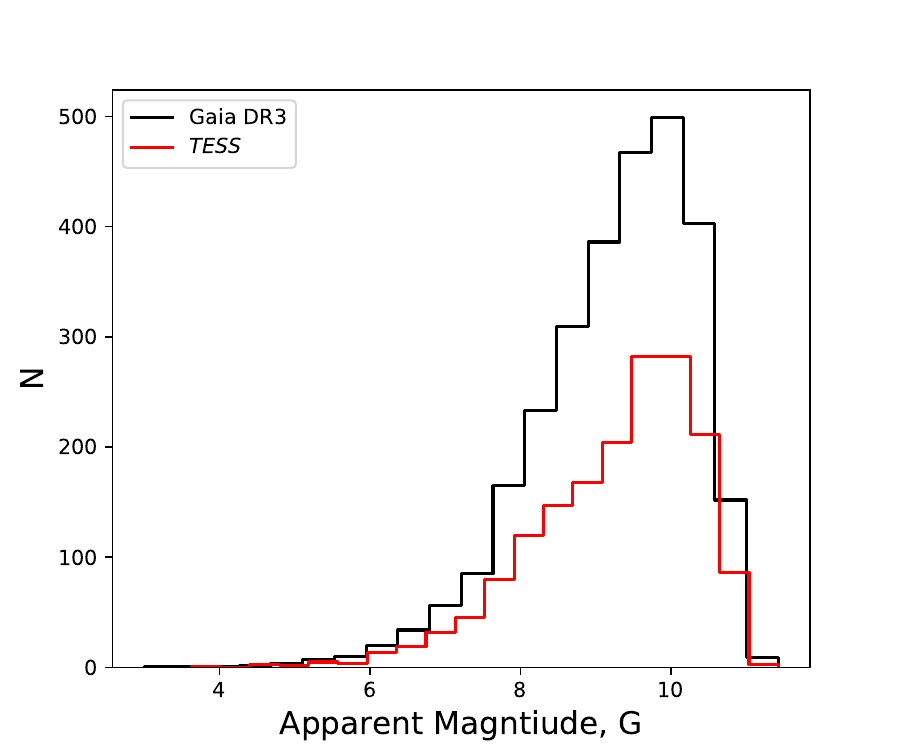}
\includegraphics[width=1.0\linewidth]{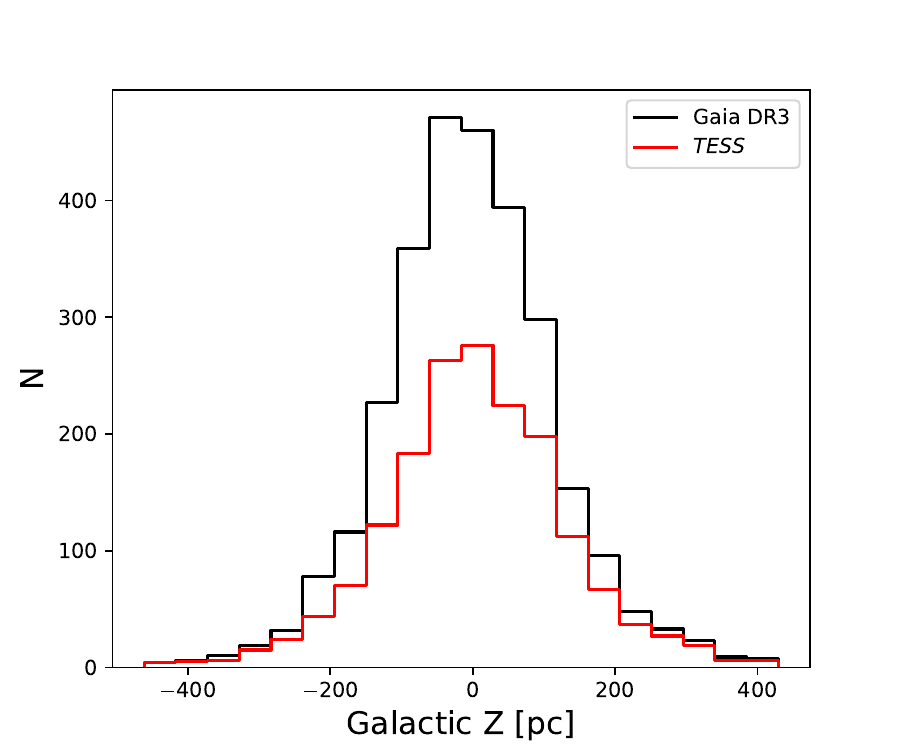}
\caption{The black histograms show the distributions of the \ngaia\ stars in our Gaia DR3 sample as a function of apparent magnitude (top) and distance from the Galactic plane (bottom).  The red histograms show the \ntess\ stars that have 10-minute \tess\ FFI light curves.
}
\label{fig:gaia_hist}
\end{figure}

\begin{figure}
\includegraphics[width=1.0\linewidth]{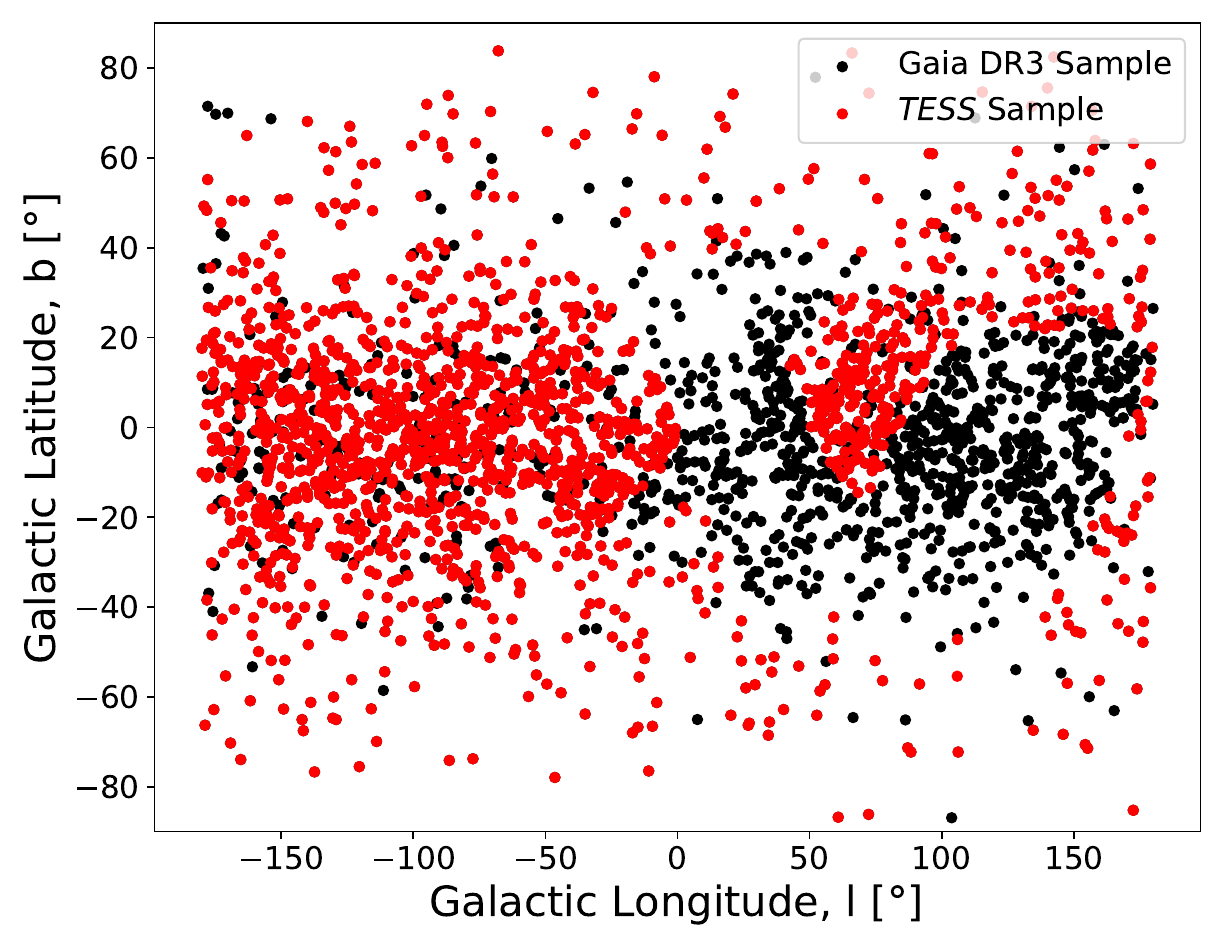}
\caption{Distribution on the sky in Galactic coordinates of the \ngaia\ Gaia DR3 stars in our sample (black points) and the \ntess\ stars with \tess\ 10-minutes light curves (red points).
}
\label{fig:galactic}
\end{figure}

Our final Gaia sample contained \ngaia\ stars. Of these, \tess\ 10-minute FFI data were available for \ntess\ stars (60\%), from the third and fourth years of the mission (Cycles 3 and 4; Sectors 27--55).  
These \ntess\ stars define the sample that we have studied in this paper.
For reference, we note that only half these stars (880 out of \ntess) had 2-minute TESS light curves, and so our use of FFIs has allowed us to obtain a much fuller sample that is essentially complete within its specified parameter space.

The upper panel of Fig.~\ref{fig:gaia_hist} shows the distribution of apparent $G$ magnitudes of the Gaia sample (\ngaia\ stars) and the final \tess\ sample (\ntess\ stars).  
The lower panel shows the Galactic $z$ coordinate, calculated using the \texttt{astropy} function \texttt{coordinates.SkyCoord}.
The exponential decline in number density with $|z|$ is due to the vertical extent of the Milky Way disk \citep[e.g.,][]{Rix+Bovy2013}.
Figure~\ref{fig:galactic} shows the distribution in Galactic sky coordinates.  The gap in \tess\ coverage reflects the fact that the spacecraft did not observe much of the ecliptic plane during Cycles~3 and~4.

\subsection{Analysis of \tess\ light curves}
\label{sec:tess}

Light curves corrected for systematic errors were extracted from the \tess\ FFIs using \texttt{eleanor} version 2.0.5 \citep{Feinstein++2019}.
For each star, we selected all sectors for which \tess\ obtained observations at 10-minute cadence. 
We produced light curves using the default \texttt{eleanor} settings and the largest possible aperture (a 5 $\times$ 5 pixel box centered on the target), which we found to maximise the signal-to-noise ratio (SNR) of \dscuti\ oscillations. 
We then selected the \texttt{eleanor} ``corrected flux'' light curves, which removed long-period systematics that correlated with telescope pointing, time, or background levels.  This served as a high-pass filter that removed variation slower than $\sim$1 day.
Among our targets, $38$\% were observed at 10-minute cadence in more than one sector, and we stitched their light curves together and treated the combined observations as a single data set.
Finally, we calculated the Fourier amplitude spectrum of each light curve up to a maximum frequency of 72\,\cd, which is the Nyquist frequency of the 10-minute \tess\ data. 

\subsection{Identification of \texorpdfstring{$\delta$}{delta}~Scuti stars}
\label{sec:class}

\begin{figure*}
    \includegraphics[width=\columnwidth]{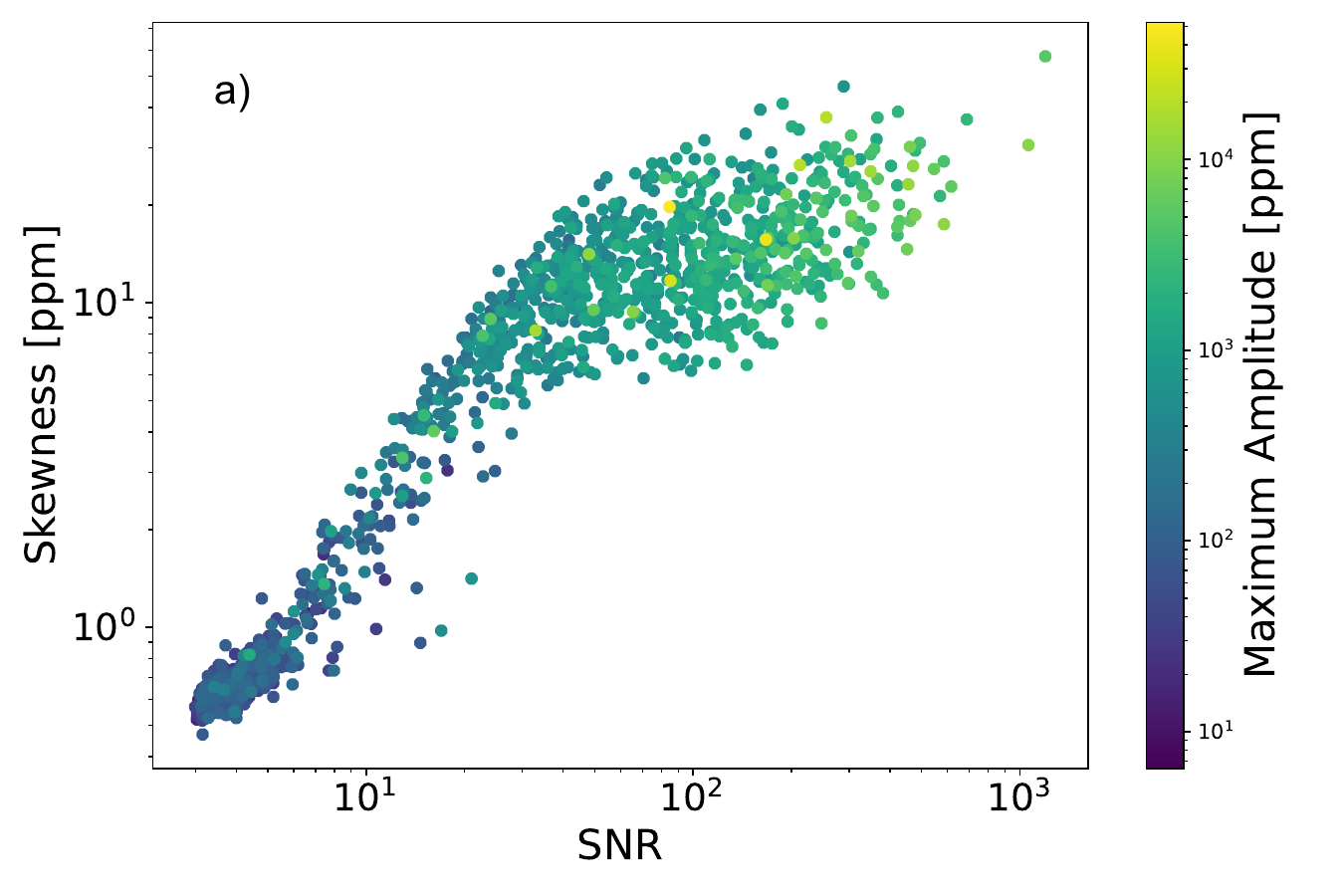}
    \includegraphics[width=0.8\columnwidth]{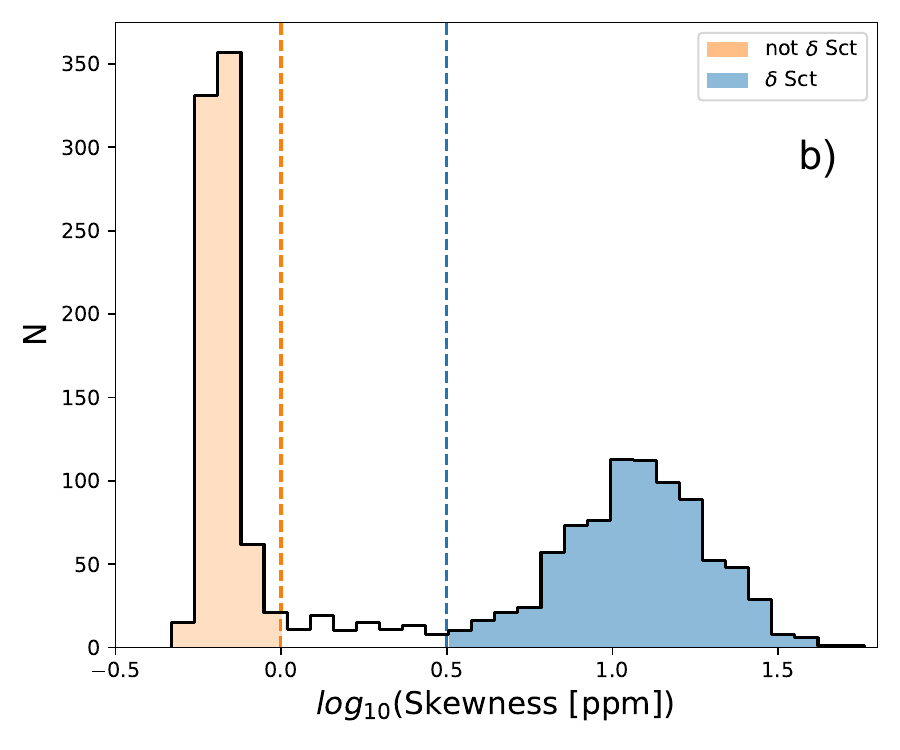}
\caption{Left: Skewness of the amplitude spectrum above 5\,\cd\ for \ntess\ stars with 10-minute \tess\ FFIs, versus the SNR of the highest peak above 5\,\cd. The colour scale depicts the amplitude of this highest peak. Right: Histogram of the skewness measurements (see Sec.~\ref{sec:tess} for details).}
\label{fig:sk}
\end{figure*}

To identify \dscuti\ pulsators, we followed a similar method to \citet{Murphy2019}. For each Fourier spectrum, we calculated the skewness of the distribution of amplitudes for frequencies above 5\,\cd . We have found this to be an excellent first step for identifying \dscuti\ stars, since their Fourier spectra have a series of strong peaks amid a relatively flat distribution of background noise. We also measured the amplitude and SNR of the highest peak above 5\,\cd, where the noise was measured as the mean amplitude at high frequencies (60--70\,\cd).  

Figure~\ref{fig:sk}a shows the results, where skewness is plotted against the SNR of the highest peak (with peak amplitudes indicated by colour). The non-\dscuti{}s are grouped at the lower left, with low skewness and with a highest peak that has low SNR and is presumably due to noise. Stars at the upper right, with a skewness above $10^{0.5}$, are mostly \dscuti\ stars.  The `neck' of stars connecting these groups, which contains about 5\% of the sample (Figure~\ref{fig:sk}b), are less clear-cut and we inspected these amplitude spectra to confirm their status.  In many cases, the peaks above 5\,\cd\ were harmonics of peaks at lower frequency that arise from other types of variability, such as gravity and Rossby modes \citep{Aerts2021}, and eclipsing or contact binaries \citep{Southworth2021}. 
We also used the period--luminosity relation that is followed by the fundamental radial mode of \dscuti{} stars \citep{Ziaali2019, Barac++2022} as a guide of where to expect the lowest-frequency p modes. In the end, we classified 59 of the 92 `neck' stars as \dscuti\ pulsators.
As a final check, we manually inspected all amplitude spectra and found a further seven stars that had been incorrectly labelled as \dscuti{}s (these were eclipsing or contact binaries).

\begin{figure*}
\includegraphics[height=10cm]{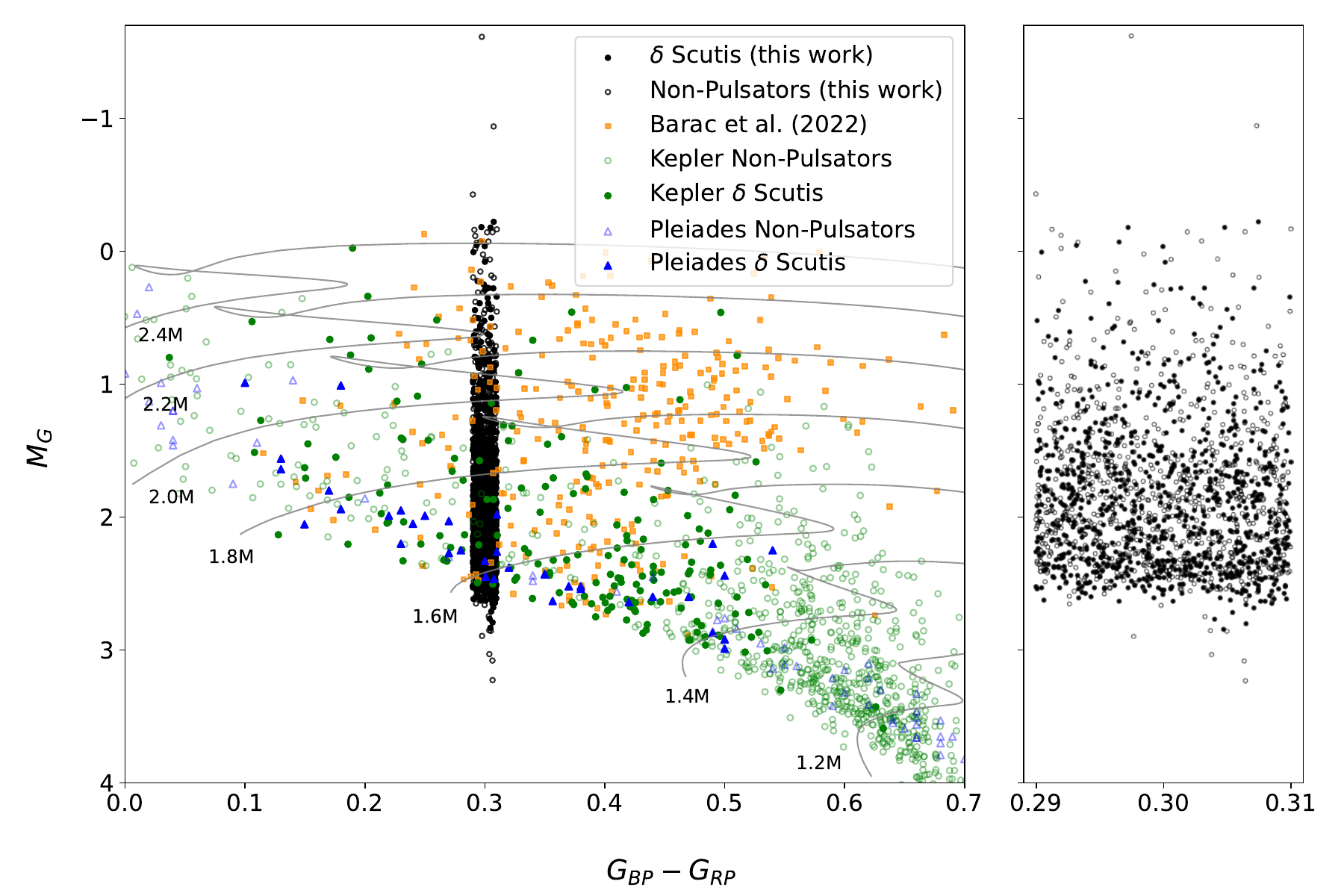}
\caption{\new{Left:} Gaia colour-magnitude showing the sample of \ntess\ stars (black circles) lying in a narrow colour strip ($0.29 < \bprp < 0.31$). Filled symbols show \dscuti\ stars and open symbols are non-\dscuti{}s. The grey lines are evolutionary tracks with solar composition \citep{Dotter2016}.
The blue triangles are stars in the Pleiades open cluster studied by \citet{Bedding2023}, the orange squares are \nbarac\ \dscuti{} stars studied by \citet{Barac++2022} that fall within 500\,pc and the green circles are the \kepler\ sample studied by \citet[][see Sec.~\ref{sec:frac}]{Murphy2019}. \new{Right: close-up showing the \ntess\ stars in our sample.}
}
\label{fig:cmd}
\end{figure*}

\begin{table}
    \centering
    \caption{List of 11 eclipsing or contact binaries that display \dscuti\ pulsations}
    \begin{tabular}{rrr}
\toprule
    TIC   & Name  &  $G$ \\
\midrule          
 56127811 & HD 29972  &  9.70 \\
 80106276 & HD 67195  &  8.58 \\
151238693 &           & 10.43 \\
163559588 & HD 100926 &  9.71 \\ 
233976224 & HD 262353 &  9.79 \\ 
248990523 &           &  9.90 \\
258351350 & HZ Dra    &  8.09 \\
279900855 & HD 292098 &  9.36 \\
306371075 & HD  13018 &  6.59 \\ 
394396270 & HD 127472 & 10.21 \\
463402815 & HD  89160 & 10.17 \\
\bottomrule
\end{tabular}
    \label{tab:eb-dsct}
\end{table}

\begin{table}
    \centering
    \caption{List of 36 eclipsing or contact binaries that do not display obvious \dscuti\ pulsations}
    \begin{tabular}{rrrr}
\toprule
    TIC &   $G$   &  TIC &   $G$ \\
\midrule
 26858469 &  9.79 &  215336287 &  9.27 \\
 33520777 &  9.94 &  222020466 &  7.17 \\
 60658382 &  6.07 &  235187292 & 10.45 \\
 67043253 &  8.45 &  264899149 & 10.34 \\
 69459754 &  9.83 &  266735682 & 10.52 \\
 72852297 &  9.42 &  268404844 &  9.05 \\
 73945470 &  8.82 &  290146287 & 10.51 \\
 74528318 &  9.64 &  300955434 &  9.77 \\
101417376 & 10.02 &  332337097 &  9.58 \\
120689840 &  9.33 &  347081956 &  8.27 \\
134907045 &  9.65 &  348830267 &  9.00 \\
147972932 &  8.60 &  352342181 &  9.91 \\
149846643 &  9.48 &  364398410 &  9.47 \\
150443185 &  9.62 &  364965655 &  8.43 \\
161840049 & 10.04 &  365065420 & 10.33 \\
172903165 & 10.11 &  367436903 & 10.36 \\
188573609 &  9.83 &  386588115 &  9.12 \\
189475510 & 10.35 &  391461666 &  8.68 \\
195612622 & 10.22 &  421285186 &  7.10 \\
196052588 & 10.04 &  451878588 & 10.76 \\
199549116 &  8.57 &  453620692 &  9.61 \\
203652249 &  9.98 &  814744585 &  9.19 \\
204318223 &  9.18 &  985576490 &  8.06 \\ 

\bottomrule
\end{tabular}

    \label{tab:eb}
\end{table}

Overall, from our sample of \ntess\ stars we identified \ndsct\ \dscuti\ pulsators, as shown in Fig.~\ref{fig:cmd} in a colour-magnitude diagram.  We note that three stars in our sample are members of the Pleiades (HD~20655, V1228~Tau and HD~23863;
blue triangles in Fig.~\ref{fig:cmd}), as recently studied by \citet{Bedding2023}. 
\new{To check the reliability of our results, we cross-matched our sample with the \nbarac\ known \dscuti\ stars studied using \tess\ 2-min light curves by \citet{Barac++2022}. Ten stars are in common and we successfully detected \dscuti\ pulsations in all of them using the 200-s FFI light curves. }
We also identified 51 eclipsing or contact binaries in our sample, of which 11 show clear \dscuti\ pulsations (Table~\ref{tab:eb-dsct}) and 46 do not (Table~\ref{tab:eb}).
Two of the stars in Table~\ref{tab:eb-dsct} were previously known to be eclipsing binaries with a pulsating component: TIC~56127811 (HD~29972; \citealt{Chen++2022, Shi++2022}) and TIC~258351350 (HZ~Dra; \citealt{Chen++2022}).




\section{Discussion}

\subsection{Amplitude spectra}
\label{sec:amp-spectra}

\begin{figure*}
\includegraphics[width=\linewidth]{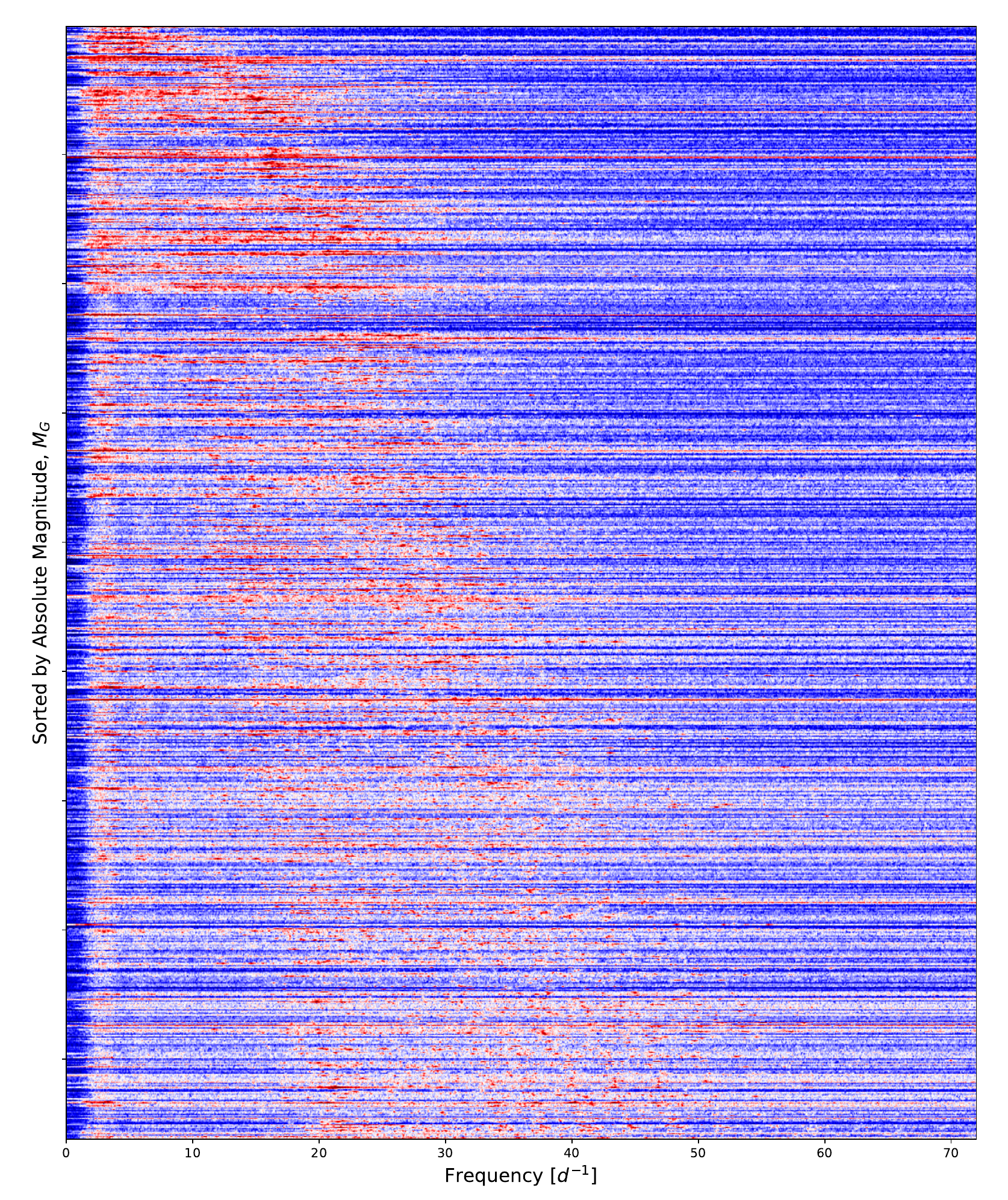}
\caption{Stacked amplitude spectra for the \ndsct\ \dscuti\ stars in our sample, sorted by absolute magnitude (brightest stars at the top of the diagram). 
}
\label{fig:stacked}
\end{figure*}

Our sample allows us to examine the properties of \dscuti\ stars within a narrow temperature band as a function of luminosity.  Figure~\ref{fig:stacked} shows the amplitude spectra of the \ndsct\ \dscuti\ stars sorted according to absolute magnitude ($M_G$), with luminosity increasing from bottom to top.  Note that the vertical stripes at 3, 6, 9 and 12\,\cd\ are artifacts caused by ringing from the high-pass filter in the \texttt{eleanor} pipeline (see Sec.~\ref{sec:tess}). Apart from the feature at 3\,\cd\ (which is below the \dscuti\ regime), these are very low-level and are barely visible in the individual amplitude spectra. 

\begin{figure*}
\includegraphics[width=\linewidth]{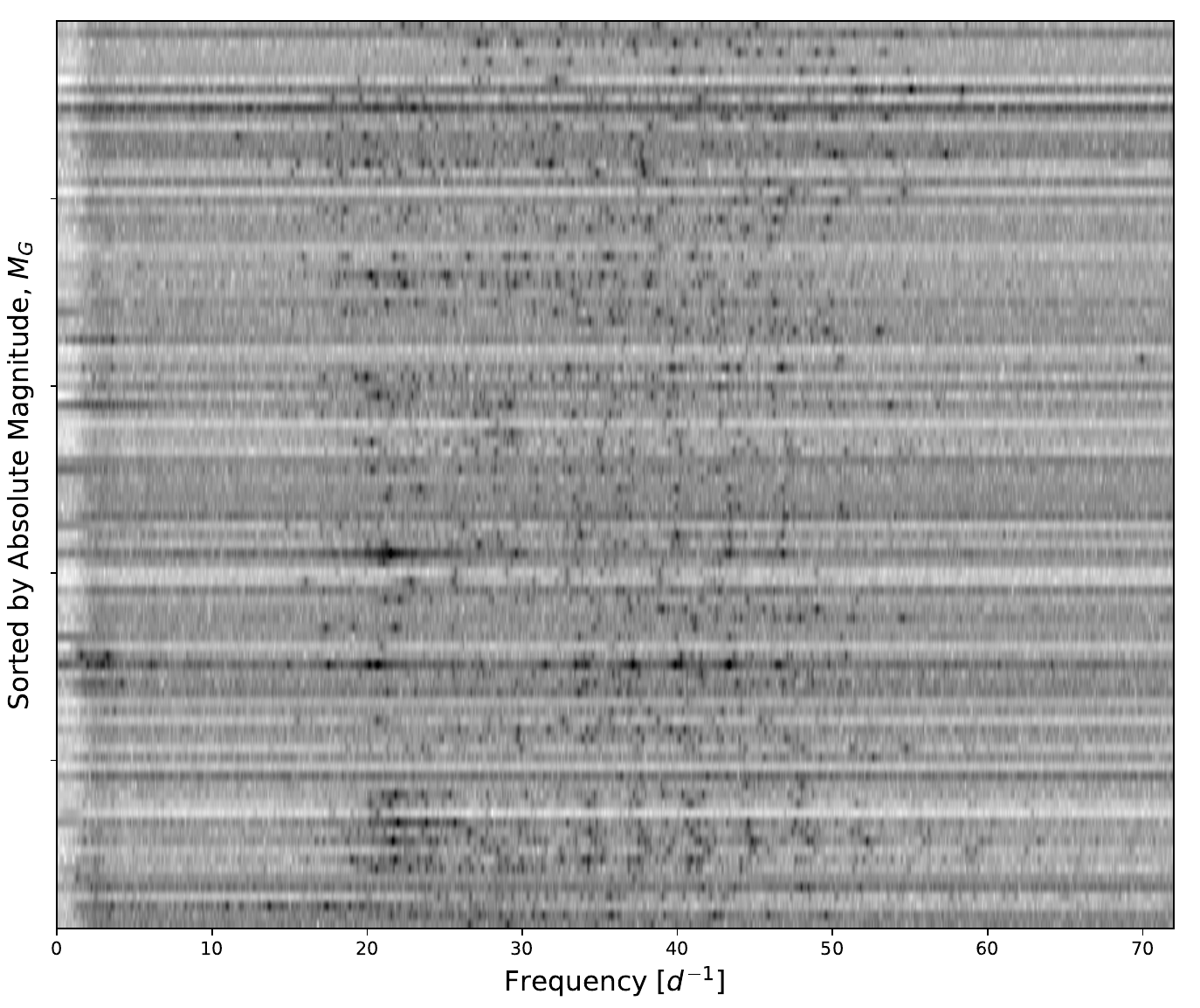}
\caption{Stacked amplitude spectra from 0--72 \cd\ (the Nyquist frequency) for the 100 least luminous \dscuti\ stars. The spectra are sorted by absolute magnitude, with the brightest stars at the top of the diagram.
}
\label{fig:stacked-bottom}
\end{figure*}

Overall, Fig.~\ref{fig:stacked} confirms that the frequencies of \dscuti\ stars decrease with increasing luminosity. This reflects the well-established property that the frequencies of p~modes scale as the square root of stellar density \citep[e.g.,][]{Aerts++2010-book}. 
Figure~\ref{fig:stacked-bottom} shows the subset of the 100 least luminous (and, therefore, predominantly the youngest) \dscuti\ stars in the sample. We clearly see the regular patterns of high-frequency modes that were seen in similar plots using data from CoRoT \citep{Michel2017} and \kepler\ \citep{Bowman+Kurtz2018}, and were studied in detail for \tess\ stars by \citet{Bedding2020}.  Some examples are discussed in Sec.~\ref{sec:high-freq}.

\subsection{Period--luminosity relation}
\label{sec:pl}

The trend in Fig.~\ref{fig:stacked} of pulsation frequency decreasing with increasing luminosity can also be visualised in the period--luminosity (P--L) diagram.  
The \dscuti\ P--L relation has previously been examined using parallaxes from Hipparcos \citep{McNamara1997,McNamara2011}, Gaia DR2 \citep{Ziaali2019,Jayasinghe2020} and Gaia DR3 \citep{Poro2021, Barac++2022, Gaia-De-Ridder++2023}, and also in the Large Magellanic Cloud \citep{Martinez-Vazquez++2022}. 

\begin{figure}
\includegraphics[width=\linewidth]{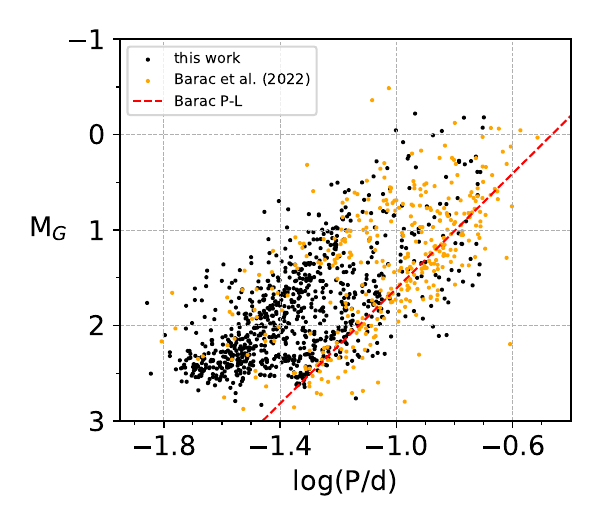}
\caption{Period--luminosity relation for our sample of \ndsct\ \dscuti\ stars (black circles) and \nbarac\ \dscuti\ stars from the ground-based catalogues \citep{Barac++2022}. The red line is the P--L relation fitted by \citet{Barac++2022}, which corresponds to the fundamental radial mode (see Sec.~\ref{sec:pl}).
}
\label{fig:pl}
\end{figure}

The P--L relation for our sample is shown in Fig.~\ref{fig:pl} (black circles), which plots the absolute magnitude of each star versus the period of its dominant pulsation mode (the strongest peak above 5\,\cd). On the same figure, the orange symbols show \nbarac\ \dscuti\ stars from ground-based catalogues (mainly \citealt{Rodriguez2000}) that were studied by \citet{Barac++2022}, using parallaxes from Gaia DR3 and periods measured from \tess\ light curves.  

The main P--L relation for \dscuti\ stars corresponds to the fundamental radial mode (red line in Fig.~\ref{fig:pl}, from \citealt{Barac++2022}).  This mode tends to be strong in higher-amplitude stars \citep[e.g.,][]{McNamara2011}, which explains why the relation is so clear in catalogues compiled using ground-based light curves. Some of the stars in our sample do fall on this relation, but the majority lie to its left, indicating that they pulsate in higher overtones. Indeed, many of them fall on a second relation that is a factor of two shorter in period than the fundamental ($-0.30$ in $\log P$). This second ridge was first identified by \citet{Ziaali2019} and studied in more detail by \citet{Jayasinghe2020} and \citet{Barac++2022}. Interestingly, we find that the dominant peak in many of our stars lies on this second relation.

\begin{figure}
\includegraphics[width=\linewidth]{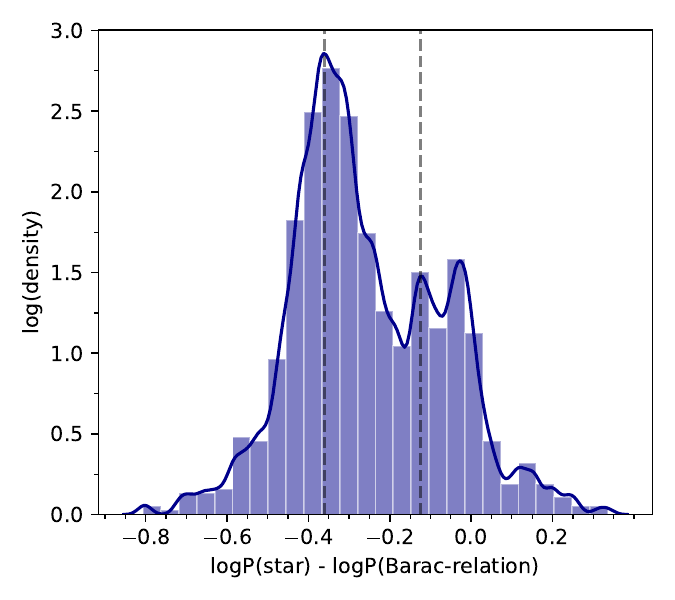}
\caption{Histogram of the distance in $\log P$ of the \ndsct\ \dscuti\ stars from the fundamental P--L relation, calculated as the horizontal distance from the red diagonal line in Fig.~\ref{fig:pl}. Overlaid is a fixed-bandwidth KDE (solid blue line), which has peaks marked by vertical dashed lines (see Sec.~\ref{sec:pl}).}
\label{fig:distancepl}
\end{figure}

Fig.~\ref{fig:distancepl} shows a histogram of the horizontal distance of the \ndsct\ \dscuti\ stars in our sample from the red diagonal line in Fig.~\ref{fig:pl}, which is the P--L relation of the fundamental mode \citep{Barac++2022}. Overlaid as a solid curve is a fixed-bandwidth KDE (kernel density estimate; \citealt{Terrell+Scott1992}).  We see a narrow excess of stars at a distance in $\log P$ of about $-0.11$ (a factor of 0.78), which we can identify with first-overtone pulsations \citep{Petersen+CD1996}.  The much larger peak at $-0.35$ corresponds to the second relation discussed above and indicates that the strongest mode in many low-amplitude \dscuti{} stars is the third or fourth overtone, rather than the fundamental or the first overtone.

\subsection{Fraction of pulsators}
\label{sec:frac}

\begin{figure}
\includegraphics[width=\linewidth]{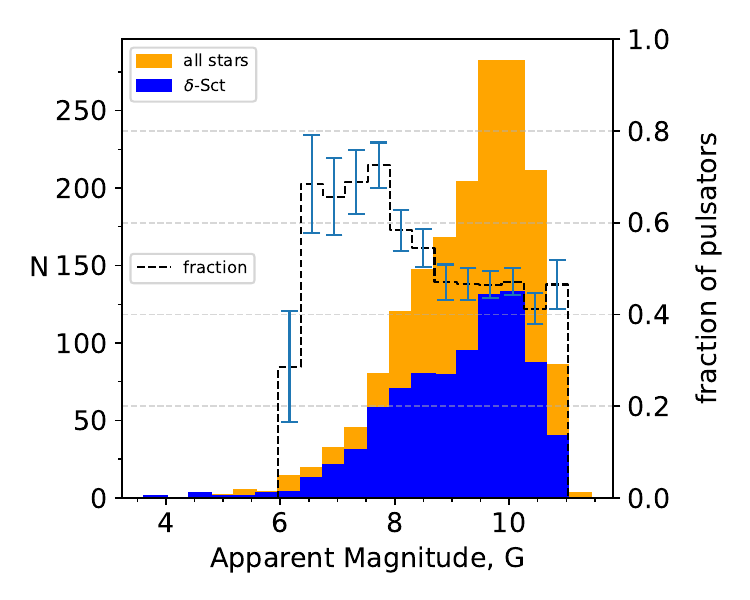}
\caption{Histograms of apparent magnitude, $G$, for \ndsct\ \dscuti\ stars (blue) out of a total sample of \ntess\ stars (orange). Referring to the right-hand axis, the fraction of \dscuti\ pulsators in each bin is shown by the dashed black histogram.
}
\label{fig:frac}
\end{figure}

The fraction of stars within the \dscuti\ instability strip that show pulsations was measured by \citet{Murphy2019} using the \kepler\ long-cadence (30-min) light curves.  They found the fraction in the central region of the strip to be 50--70\%.
Fig.~\ref{fig:frac} shows the \ndsct\ \dscuti\ pulsators in our sample (blue histogram) and the total of \ntess\ stars (orange histogram) as a function of apparent $G$ magnitude. The fraction of pulsators (dashed black histogram) is about 70\% for the brightest stars ($G<8$), consistent with the \kepler\ result, but drops  off for fainter stars. This implies that we are missing some pulsating stars at the fainter end of our \tess\ sample.

\begin{figure*}
\includegraphics[width=0.8\linewidth]{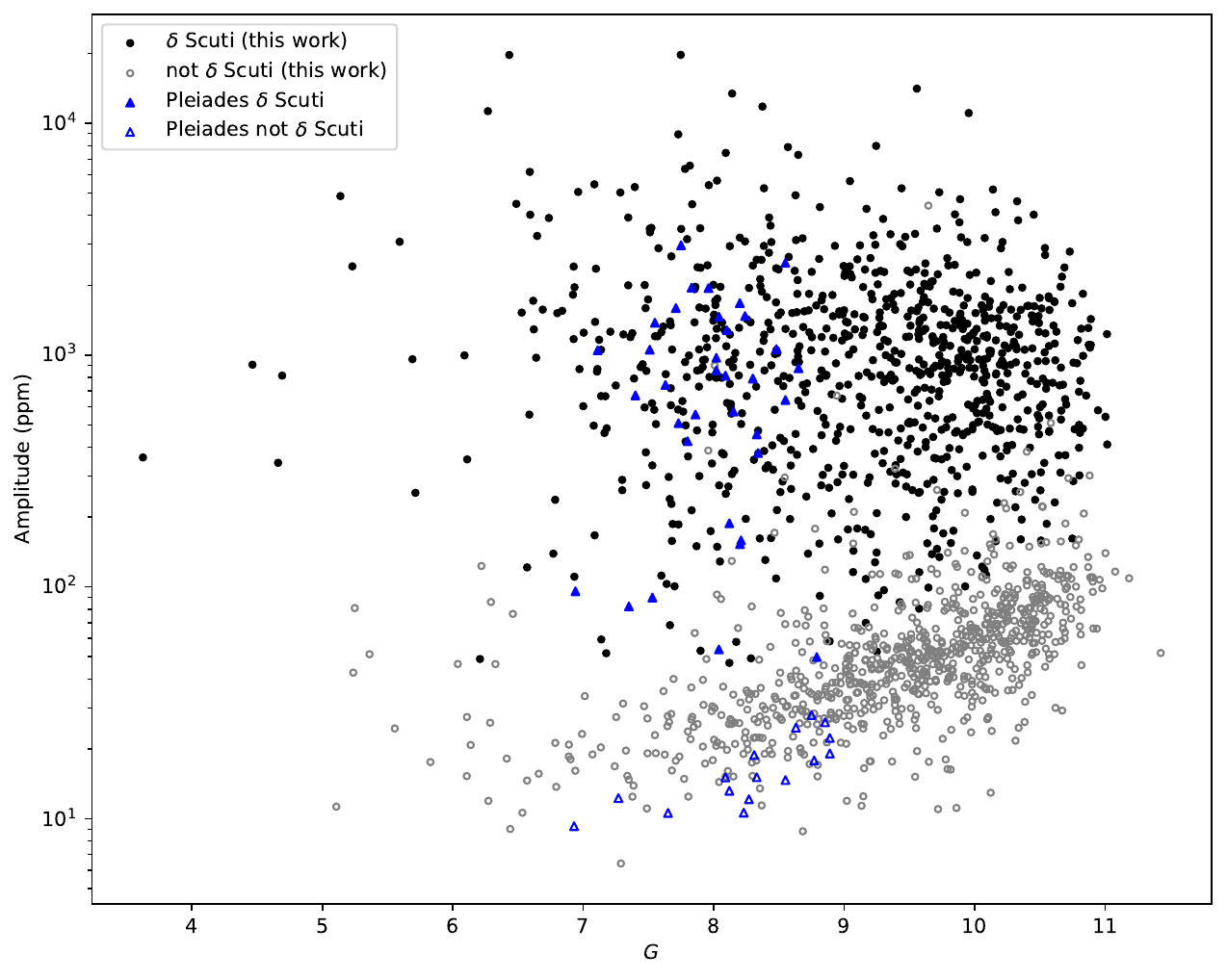}
\caption{Amplitude of the highest peak in the Fourier spectrum above 5\,\cd, plotted against apparent magnitude, $G$. }
\label{fig:amp_vs_G}
\end{figure*}

To investigate the completeness of our detections, Fig.~\ref{fig:amp_vs_G} shows the amplitude of the highest peak in the Fourier spectrum of each star (above 5\,\cd), plotted as a function of apparent $G$ magnitude.  In the non-\dscuti\ stars (open grey circles) this is a measure of the white noise, which is why we see a correlation with apparent magnitude.  For the \dscuti\ stars in our sample (filled black circles), most have amplitudes in the range 300--3000\,ppm.\footnote{The open symbol in the upper right of Fig.~\ref{fig:amp_vs_G} is from the star TIC~395181620 (HD~118476), whose light curve is contaminated by a nearby RR Lyrae (TIC 395181648; \citealt{Clementini++2019}).}
The lowest amplitudes are around 50\,ppm, but these are only detected in stars brighter than about $G=9$.  We can see from the distribution of points that we are probably missing some pulsators among the fainter stars ($G>8$), as implied by the aforementioned drop in pulsator fraction in Fig.~\ref{fig:frac}.

In Fig.~\ref{fig:amp_vs_G} we also show the Pleiades sample studied by \citet{Bedding2023}, restricted to the colour range in which pulsations occur ($0.1 < \bprp < 0.55$). The non-\dscuti\ stars (open blue triangles) have somewhat lower white-noise levels than our sample because the Pleiades was observed for three \tess\ sectors, whereas most of our sample have only one sector (keeping in mind we only included 10-min FFI data). The \dscuti\ pulsators in the Pleiades (filled blue triangles) have amplitudes down to 50\,ppm. In this case we can be fairly confident the Pleiades \dscuti\ sample is complete at this amplitude level because (i)~the Pleiades stars are all brighter than $G=9$, and (ii)~the white noise level is lower (see above).


\subsection{High-frequency \dscuti\ stars}
\label{sec:high-freq}

\begin{figure*}
\includegraphics[width=0.24\linewidth]{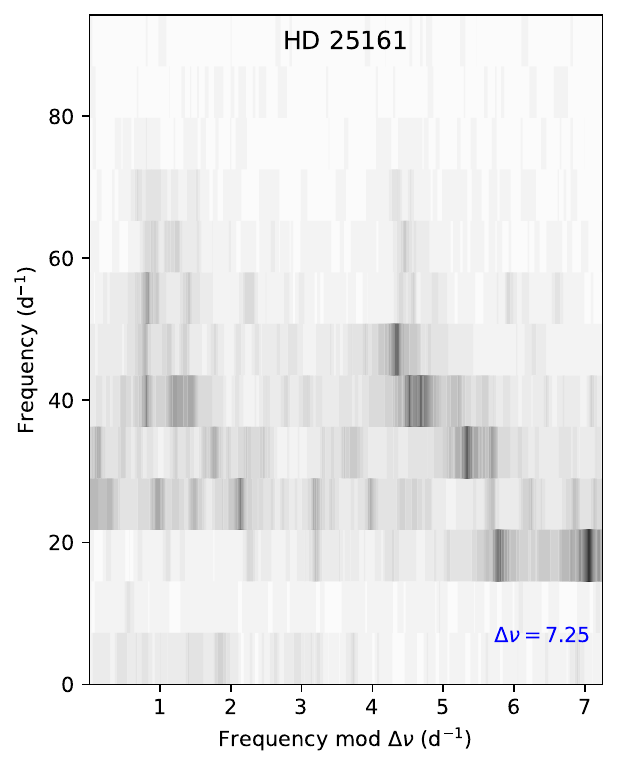}
\includegraphics[width=0.24\linewidth]{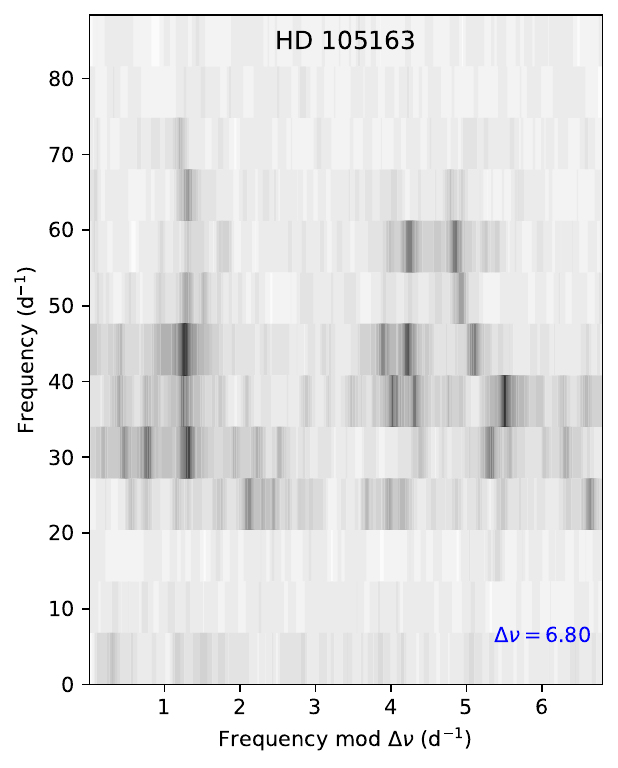}
\includegraphics[width=0.24\linewidth]{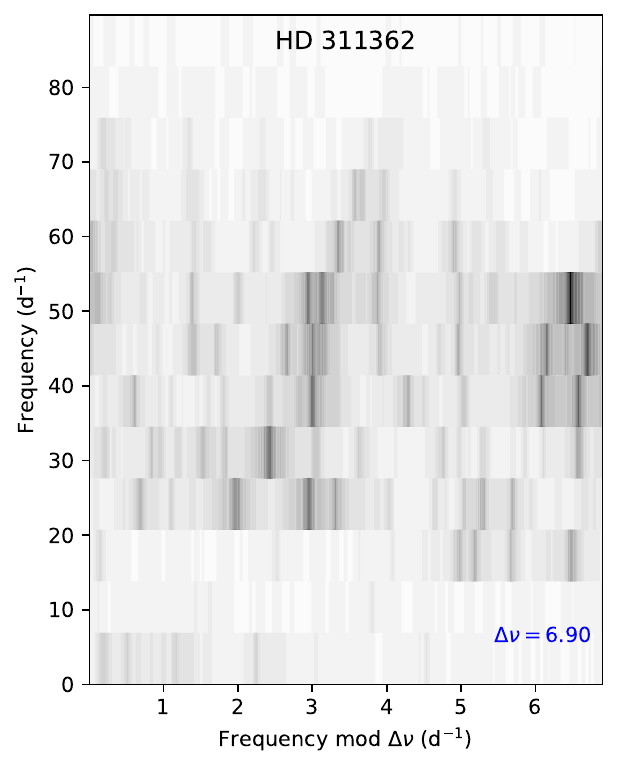}
\includegraphics[width=0.24\linewidth]{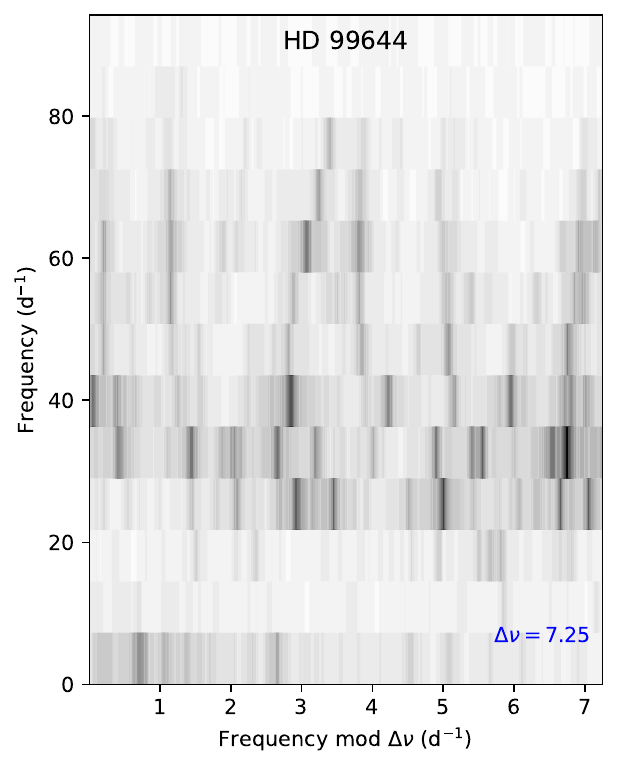}
\caption{Amplitude spectra of four newly discovered high-frequency \dscuti\ stars, shown in \echelle\ format \citep[see][]{Bedding2020}.
}
\label{fig:echelles}
\end{figure*}

As mentioned in Sec.~\ref{sec:amp-spectra}, our sample includes several \dscuti\ stars with regular patterns of high-frequency modes (see Fig.~\ref{fig:stacked-bottom}). Nine of these are included in the sample of \citet{Bedding2020}, namely  
HD~3622,
HD~24975,
HD~59594 (V349~Pup),
HD~17341,
HD~46722,
HD~31640,
HD~29783,
HD~20203,
and HD~55863.
%
%
In addition, we have identified several new examples. Four with particularly regular oscillation patterns are shown in Fig.~\ref{fig:echelles}:
\begin{itemize}
\item 
TIC 14254276 (HD 25161), which is a member of the Taurus Association \citep{Gagne++2018}.
 
\item 
TIC 86263305 (HD 105163), which does not appear to be a member of a moving group or association. \new{We have obtained spectra of this star with Minerva Australis \citep{Addison++2019}. From the 16 spectra taken on 9 nights, there is no evidence of a companion, either in radial velocities or in the line profiles. The spectra indicate a low $v\sin i$ of $11\pm7$\,km\,s$^{-1}$, and an analysis of 84 Fe lines using iSpec \citep{blancoetal2014,blanco2019} shows $T_{\rm eff} = 7600\pm200$\,K and [Fe/H] = $-0.65\pm0.26$.}

\item 
TIC 307035635 (HD 311362), which is in \tess's southern continuous viewing zone and does not appear to be a member of a moving group or association

\item 
TIC 295882266 (HD 99644), which is a member of the Lower Centaurus-Crux (LCC) association \citep{Gagne++2018}.

\end{itemize}
These are all good targets for further study and detailed modelling \new{to determine stellar ages, by taking advantage of the fact that their regular patterns allow the modes to be identified \citep[e.g.,][]{Pamos-Ortega++2022, Pamos-Ortega++2023, Steindl++2022, Murphy++2023, Palakkatharappil+Creevey2023, Scutt++2023, Panda++2024}}.

\subsection{Notes on individual stars}

\subsubsection{\texorpdfstring{$\lambda$}{lambda} Muscae: a new bright \dscuti\ star}

\begin{figure}
\includegraphics[width=\linewidth]{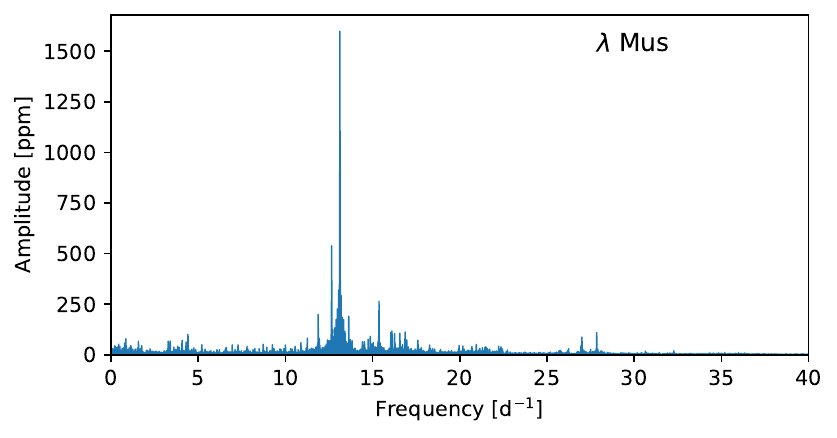}
\caption{Amplitude spectrum of the bright \dscuti\ star $\lambda$~Mus, from the \tess\ 2-min light curve (Sectors 10, 11, 37 and 38). This star was not previously known to be a \dscuti.
}
\label{fig:lambda-mus}
\end{figure}

The star $\lambda$~Mus has not previously been reported as variable but the \tess\ data clearly shows \dscuti\ pulsations (Fig.~\ref{fig:lambda-mus}). 
Its apparent magnitude of $V=3.65$ ($G=3.62$) makes $\lambda$~Mus the fourth-brightest known \dscuti, behind 
$\alpha$~Aql (Altair; $V=0.76$; \citealt{Buzasi++2005}), 
$\rho$~Pup ($V=2.81$; \citealt{Eggen1956}), and
$\gamma$~Boo ($V=3.02$; \citealt{Guthnick+Fischer1940}).
Based on its absolute magnitude, the fundamental radial mode of $\lambda$~Mus should be at about 4.5\,\cd, and so the strongest mode at 13.12\,\cd\ is presumably a higher overtone.

\subsubsection{\texorpdfstring{$\kappa^2$}{kappa2} Bootis: a bright evolved \dscuti\ star}

\begin{figure}
\includegraphics[width=\linewidth]{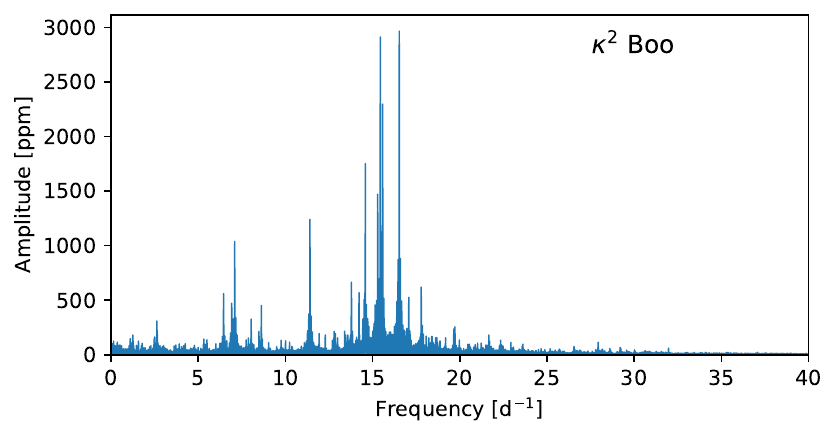}
\caption{Amplitude spectrum of the bright \dscuti\ star $\kappa^2$~Boo, from the \tess\ 2-min light curve (Sectors 22, 23, 49 and 50).  The pulsations in this star were previously studied by \citet{Frandsen++1995}.
}
\label{fig:kappa2-boo}
\end{figure}

The star $\kappa^2$~Boo is the second brightest \dscuti\ pulsator in our sample, and was studied in detail by \citet{Frandsen++1995}. The \tess\ data confirm the pulsations (Fig.~\ref{fig:kappa2-boo}).
Based on its absolute magnitude, the fundamental radial mode of $\kappa^2$~Boo should be at about 5.8\,\cd, and so the strongest modes at 15--16\,\cd\ must be overtones.

\section{Conclusions}

We used Gaia DR3 to select a sample of \ngaia\ stars within 500\,pc of the Sun that lie in a narrow colour range in the centre of the \dscuti\ instability strip ($0.29 < \bprp < 0.31$).  For \ntess\ of these stars, \tess\ 10-minute Full Frame Images (Sectors 27--55) are available and we used the {\tt eleanor} package \citep{Feinstein++2019} to extract light curves.  Using Fourier amplitude spectra, we identified \ndsct\ \dscuti\ stars, as well as 47 eclipsing or contact binaries (Tables~\ref{tab:eb-dsct} and~\ref{tab:eb}).

The pulsation frequencies of the \dscuti\ stars decrease with increasing luminosity (Fig.~\ref{fig:stacked}), which reflects the well-established property that the frequencies of pressure modes scale as the square root of stellar density. When we consider the strongest mode in each star, some fall on the period--luminosity relation of the fundamental radial mode but many correspond to overtones that are approximately a factor of two higher in frequency (Figs.~\ref{fig:pl} and \ref{fig:distancepl}). Many of the low-luminosity \dscuti\ stars show a series of high-frequency modes (Fig.~\ref{fig:stacked-bottom}), including four new examples with very regular spacings (Fig.~\ref{fig:echelles}). 

The fraction of stars in our sample that show \dscuti\ pulsations (Fig.~\ref{fig:frac}) is about 70\% for the brightest stars ($G<8$), consistent with results from \kepler\ \citep{Murphy2019}. However, for fainter stars the fraction drops to about 45\%, indicating that we are missing some pulsating stars at the fainter end. This is confirmed by Fig.~\ref{fig:amp_vs_G}, which shows that a single sector of \tess\ data only detects the lowest-amplitude \dscuti\ pulsations (around 50\,ppm) in stars down to about $G=9$.

Overall, these results confirm the power of \tess\ and Gaia for studying pulsating stars. We plan to expand this work to encompass the entire \tess\ data set across the whole instability strip.

\section*{Acknowledgements}

We gratefully acknowledge support from the Australian Research Council through Future Fellowship FT210100485, and Laureate Fellowship FL220100117.
This work has made use of data from the European Space Agency (ESA) mission {\em Gaia}, (\url{https://www.cosmos.esa.int/gaia}), 
processed by the {\em Gaia} Data Processing and Analysis Consortium (DPAC, \url{https://www.cosmos.esa.int/web/gaia/dpac/consortium}). Funding for the DPAC has been provided by national institutions, in particular the institutions participating in the {\em Gaia} Multilateral Agreement.
We are grateful to the entire Gaia and \tess\ teams for providing the data used in this paper.
This work made use of several publicly available {\tt python} packages: {\tt astropy} \citep{astropy:2013,astropy:2018}, 
{\tt lightkurve} \citep{lightkurve2018},
{\tt matplotlib} \citep{matplotlib2007}, 
{\tt numpy} \citep{numpy2020}, and 
{\tt scipy} \citep{scipy2020}.

\section*{Data Availability}

The \tess\ data underlying this article are available at the MAST Portal (Barbara A. Mikulski Archive for Space Telescopes), at \url{https://mast.stsci.edu/portal/Mashup/Clients/Mast/Portal.html}
\ifarxiv
    \input{output.bbl} 
\else
    \bibliographystyle{mnras}
    \bibliography{references}

\begin{thebibliography}{}
\makeatletter
\relax
\def\mn@urlcharsother{\let\do\@makeother \do\$\do\&\do\#\do\^\do\_\do\%\do\~}
\def\mn@doi{\begingroup\mn@urlcharsother \@ifnextchar [ {\mn@doi@}
  {\mn@doi@[]}}
\def\mn@doi@[#1]#2{\def\@tempa{#1}\ifx\@tempa\@empty \href
  {http://dx.doi.org/#2} {doi:#2}\else \href {http://dx.doi.org/#2} {#1}\fi
  \endgroup}
\def\mn@eprint#1#2{\mn@eprint@#1:#2::\@nil}
\def\mn@eprint@arXiv#1{\href {http://arxiv.org/abs/#1} {{\tt arXiv:#1}}}
\def\mn@eprint@dblp#1{\href {http://dblp.uni-trier.de/rec/bibtex/#1.xml}
  {dblp:#1}}
\def\mn@eprint@#1:#2:#3:#4\@nil{\def\@tempa {#1}\def\@tempb {#2}\def\@tempc
  {#3}\ifx \@tempc \@empty \let \@tempc \@tempb \let \@tempb \@tempa \fi \ifx
  \@tempb \@empty \def\@tempb {arXiv}\fi \@ifundefined
  {mn@eprint@\@tempb}{\@tempb:\@tempc}{\expandafter \expandafter \csname
  mn@eprint@\@tempb\endcsname \expandafter{\@tempc}}}

\bibitem[\protect\citeauthoryear{{Addison} et~al.,}{{Addison}
  et~al.}{2019}]{Addison++2019}
{Addison} B.,  et~al., 2019, \mn@doi [\pasp] {10.1088/1538-3873/ab03aa}, \href
  {https://ui.adsabs.harvard.edu/abs/2019PASP..131k5003A} {131, 115003}

\bibitem[\protect\citeauthoryear{{Aerts}}{{Aerts}}{2021}]{Aerts2021}
{Aerts} C.,  2021, \mn@doi [Reviews of Modern Physics]
  {10.1103/RevModPhys.93.015001}, \href
  {https://ui.adsabs.harvard.edu/abs/2021RvMP...93a5001A} {93, 015001}

\bibitem[\protect\citeauthoryear{{Aerts}, {Christensen-Dalsgaard}  \&
  {Kurtz}}{{Aerts} et~al.}{2010}]{Aerts++2010-book}
{Aerts} C.,  {Christensen-Dalsgaard} J.,   {Kurtz} D.~W.,  2010,
  {Asteroseismology}.
Springer

\bibitem[\protect\citeauthoryear{{Antoci} et~al.,}{{Antoci}
  et~al.}{2019}]{Antoci++2019}
{Antoci} V.,  et~al., 2019, \mn@doi [\mnras] {10.1093/mnras/stz2787}, \href
  {https://ui.adsabs.harvard.edu/abs/2019MNRAS.490.4040A} {490, 4040}

\bibitem[\protect\citeauthoryear{{Astropy Collaboration}}{{Astropy
  Collaboration}}{2013}]{astropy:2013}
{Astropy Collaboration} 2013, \mn@doi [\aap] {10.1051/0004-6361/201322068},
  \href {http://adsabs.harvard.edu/abs/2013A%26A...558A..33A} {558, A33}

\bibitem[\protect\citeauthoryear{{Astropy Collaboration}}{{Astropy
  Collaboration}}{2018}]{astropy:2018}
{Astropy Collaboration} 2018, \mn@doi [\aj] {10.3847/1538-3881/aabc4f}, \href
  {https://ui.adsabs.harvard.edu/abs/2018AJ....156..123A} {156, 123}

\bibitem[\protect\citeauthoryear{{Balona} \& {Ozuyar}}{{Balona} \&
  {Ozuyar}}{2020}]{Balona+Ozuyar2020}
{Balona} L.~A.,  {Ozuyar} D.,  2020, \mn@doi [\mnras] {10.1093/mnras/staa670},
  \href {https://ui.adsabs.harvard.edu/abs/2020MNRAS.493.5871B} {493, 5871}

\bibitem[\protect\citeauthoryear{{Balona}, {Holdsworth}  \& {Cunha}}{{Balona}
  et~al.}{2019}]{Balona++2019}
{Balona} L.~A.,  {Holdsworth} D.~L.,   {Cunha} M.~S.,  2019, \mn@doi [\mnras]
  {10.1093/mnras/stz1423}, \href
  {https://ui.adsabs.harvard.edu/abs/2019MNRAS.487.2117B} {487, 2117}

\bibitem[\protect\citeauthoryear{{Barac}, {Bedding}, {Murphy}  \&
  {Hey}}{{Barac} et~al.}{2022}]{Barac++2022}
{Barac} N.,  {Bedding} T.~R.,  {Murphy} S.~J.,   {Hey} D.~R.,  2022, \mn@doi
  [\mnras] {10.1093/mnras/stac2132}, \href
  {https://ui.adsabs.harvard.edu/abs/2022MNRAS.516.2080B} {516, 2080}

\bibitem[\protect\citeauthoryear{{Bedding} et~al.,}{{Bedding}
  et~al.}{2020}]{Bedding2020}
{Bedding} T.~R.,  et~al., 2020, \mn@doi [\nat] {10.1038/s41586-020-2226-8},
  \href {https://ui.adsabs.harvard.edu/abs/2020Natur.581..147B} {581, 147}

\bibitem[\protect\citeauthoryear{{Bedding} et~al.,}{{Bedding}
  et~al.}{2023}]{Bedding2023}
{Bedding} T.~R.,  et~al., 2023, \mn@doi [\apjl] {10.3847/2041-8213/acc17a},
  \href {https://ui.adsabs.harvard.edu/abs/2023ApJ...946L..10B} {946, L10}

\bibitem[\protect\citeauthoryear{{Blanco-Cuaresma}}{{Blanco-Cuaresma}}{2019}]{blanco2019}
{Blanco-Cuaresma} S.,  2019, \mn@doi [\mnras] {10.1093/mnras/stz549}, \href
  {https://ui.adsabs.harvard.edu/abs/2019MNRAS.486.2075B} {486, 2075}

\bibitem[\protect\citeauthoryear{{Blanco-Cuaresma}, {Soubiran}, {Heiter}  \&
  {Jofr{\'e}}}{{Blanco-Cuaresma} et~al.}{2014}]{blancoetal2014}
{Blanco-Cuaresma} S.,  {Soubiran} C.,  {Heiter} U.,   {Jofr{\'e}} P.,  2014,
  \mn@doi [\aap] {10.1051/0004-6361/201423945}, \href
  {https://ui.adsabs.harvard.edu/abs/2014A&A...569A.111B} {569, A111}

\bibitem[\protect\citeauthoryear{{Bowman} \& {Kurtz}}{{Bowman} \&
  {Kurtz}}{2018}]{Bowman+Kurtz2018}
{Bowman} D.~M.,  {Kurtz} D.~W.,  2018, \mnras, \href
  {http://adsabs.harvard.edu/abs/2018MNRAS.476.3169B} {476, 3169}

\bibitem[\protect\citeauthoryear{{Buzasi} et~al.,}{{Buzasi}
  et~al.}{2005}]{Buzasi++2005}
{Buzasi} D.~L.,  et~al., 2005, \mn@doi [\apj] {10.1086/426704}, \href
  {https://ui.adsabs.harvard.edu/#abs/2005ApJ...619.1072B} {619, 1072}

\bibitem[\protect\citeauthoryear{{Chen} et~al.,}{{Chen}
  et~al.}{2022}]{Chen++2022}
{Chen} X.,  et~al., 2022, \mn@doi [\apjs] {10.3847/1538-4365/aca284}, \href
  {https://ui.adsabs.harvard.edu/abs/2022ApJS..263...34C} {263, 34}

\bibitem[\protect\citeauthoryear{{Clementini} et~al.,}{{Clementini}
  et~al.}{2019}]{Clementini++2019}
{Clementini} G.,  et~al., 2019, \mn@doi [\aap] {10.1051/0004-6361/201833374},
  \href {https://ui.adsabs.harvard.edu/abs/2019A&A...622A..60C} {622, A60}

\bibitem[\protect\citeauthoryear{{Daszy{\'n}ska-Daszkiewicz}, {Walczak},
  {Pamyatnykh}, {Szewczuk}  \& {Niewiadomski}}{{Daszy{\'n}ska-Daszkiewicz}
  et~al.}{2023}]{Daszyska-Daszkiewicz++2023}
{Daszy{\'n}ska-Daszkiewicz} J.,  {Walczak} P.,  {Pamyatnykh} A.,  {Szewczuk}
  W.,   {Niewiadomski} W.,  2023, \mn@doi [\apjl] {10.3847/2041-8213/acade2},
  \href {https://ui.adsabs.harvard.edu/abs/2023ApJ...942L..38D} {942, L38}

\bibitem[\protect\citeauthoryear{{Dotter}}{{Dotter}}{2016}]{Dotter2016}
{Dotter} A.,  2016, \mn@doi [\apjs] {10.3847/0067-0049/222/1/8}, \href
  {https://ui.adsabs.harvard.edu/abs/2016ApJS..222....8D} {222, 8}

\bibitem[\protect\citeauthoryear{{Eggen}}{{Eggen}}{1956}]{Eggen1956}
{Eggen} O.~J.,  1956, \mn@doi [\pasp] {10.1086/126922}, \href
  {https://ui.adsabs.harvard.edu/abs/1956PASP...68..238E} {68, 238}

\bibitem[\protect\citeauthoryear{{El-Badry}, {Conroy}, {Fuller}, {Kiman}, {van
  Roestel}, {Rodriguez}  \& {Burdge}}{{El-Badry} et~al.}{2022}]{El-Badry++2022}
{El-Badry} K.,  {Conroy} C.,  {Fuller} J.,  {Kiman} R.,  {van Roestel} J.,
  {Rodriguez} A.~C.,   {Burdge} K.~B.,  2022, \mn@doi [\mnras]
  {10.1093/mnras/stac2945}, \href
  {https://ui.adsabs.harvard.edu/abs/2022MNRAS.517.4916E} {517, 4916}

\bibitem[\protect\citeauthoryear{{Feinstein} et~al.,}{{Feinstein}
  et~al.}{2019}]{Feinstein++2019}
{Feinstein} A.~D.,  et~al., 2019, \mn@doi [\pasp] {10.1088/1538-3873/ab291c},
  \href {https://ui.adsabs.harvard.edu/abs/2019PASP..131i4502F} {131, 094502}

\bibitem[\protect\citeauthoryear{{Frandsen}, {Jones}, {Kjeldsen}, {Viskum},
  {Hjorth}, {Andersen}  \& {Thomsen}}{{Frandsen} et~al.}{1995}]{Frandsen++1995}
{Frandsen} S.,  {Jones} A.,  {Kjeldsen} H.,  {Viskum} M.,  {Hjorth} J.,
  {Andersen} N.~H.,   {Thomsen} B.,  1995, \aap, \href
  {https://ui.adsabs.harvard.edu/abs/1995A&A...301..123F} {301, 123}

\bibitem[\protect\citeauthoryear{{Gagn{\'e}}, {Roy-Loubier}, {Faherty}, {Doyon}
   \& {Malo}}{{Gagn{\'e}} et~al.}{2018}]{Gagne++2018}
{Gagn{\'e}} J.,  {Roy-Loubier} O.,  {Faherty} J.~K.,  {Doyon} R.,   {Malo} L.,
  2018, \mn@doi [\apj] {10.3847/1538-4357/aac2b8}, \href
  {https://ui.adsabs.harvard.edu/abs/2018ApJ...860...43G} {860, 43}

\bibitem[\protect\citeauthoryear{{Gaia Collaboration}}{{Gaia
  Collaboration}}{2021}]{Gaia++2021}
{Gaia Collaboration} 2021, \mn@doi [\aap] {10.1051/0004-6361/202039657}, \href
  {https://ui.adsabs.harvard.edu/abs/2021A&A...649A...1G} {649, A1}

\bibitem[\protect\citeauthoryear{{Gaia Collaboration} et~al.,}{{Gaia
  Collaboration} et~al.}{2023}]{Gaia-De-Ridder++2023}
{Gaia Collaboration} et~al., 2023, \mn@doi [\aap]
  {10.1051/0004-6361/202243767}, \href
  {https://ui.adsabs.harvard.edu/abs/2023A&A...674A..36G} {674, A36}

\bibitem[\protect\citeauthoryear{{Guthnick} \& {Fischer}}{{Guthnick} \&
  {Fischer}}{1940}]{Guthnick+Fischer1940}
{Guthnick} P.,  {Fischer} H.,  1940, \mn@doi [Astronomische Nachrichten]
  {10.1002/asna.19402710205}, \href
  {https://ui.adsabs.harvard.edu/abs/1940AN....271...81G} {271, 81}

\bibitem[\protect\citeauthoryear{Harris et~al.,}{Harris
  et~al.}{2020}]{numpy2020}
Harris C.~R.,  et~al., 2020, \mn@doi [Nature] {10.1038/s41586-020-2649-2}, 585,
  357

\bibitem[\protect\citeauthoryear{Hunter}{Hunter}{2007}]{matplotlib2007}
Hunter J.~D.,  2007, Computing in Science \& Engineering, 9, 90

\bibitem[\protect\citeauthoryear{{Jayasinghe} et~al.,}{{Jayasinghe}
  et~al.}{2020}]{Jayasinghe2020}
{Jayasinghe} T.,  et~al., 2020, \mn@doi [\mnras] {10.1093/mnras/staa499}, \href
  {https://ui.adsabs.harvard.edu/abs/2020MNRAS.493.4186J} {493, 4186}

\bibitem[\protect\citeauthoryear{{Li} et~al.,}{{Li}
  et~al.}{2023}]{Li-Gang++2023}
{Li} G.,  et~al., 2023, \mn@doi [arXiv e-prints] {10.48550/arXiv.2311.16991},
  \href {https://ui.adsabs.harvard.edu/abs/2023arXiv231116991L} {p.
  arXiv:2311.16991}

\bibitem[\protect\citeauthoryear{{Lightkurve Collaboration}
  et~al.,}{{Lightkurve Collaboration} et~al.}{2018}]{lightkurve2018}
{Lightkurve Collaboration} et~al., 2018, {Lightkurve: Kepler and TESS time
  series analysis in Python}, Astrophysics Source Code Library (\mn@eprint
  {ascl} {1812.013})

\bibitem[\protect\citeauthoryear{{Mart{\'\i}nez-V{\'a}zquez}, {Salinas},
  {Vivas}  \& {Catelan}}{{Mart{\'\i}nez-V{\'a}zquez}
  et~al.}{2022}]{Martinez-Vazquez++2022}
{Mart{\'\i}nez-V{\'a}zquez} C.~E.,  {Salinas} R.,  {Vivas} A.~K.,   {Catelan}
  M.,  2022, \mn@doi [\apjl] {10.3847/2041-8213/ac9f38}, \href
  {https://ui.adsabs.harvard.edu/abs/2022ApJ...940L..25M} {940, L25}

\bibitem[\protect\citeauthoryear{{McNamara}}{{McNamara}}{1997}]{McNamara1997}
{McNamara} D.,  1997, \mn@doi [\pasp] {10.1086/133999}, \href
  {http://adsabs.harvard.edu/abs/1997PASP..109.1221M} {109, 1221}

\bibitem[\protect\citeauthoryear{{McNamara}}{{McNamara}}{2011}]{McNamara2011}
{McNamara} D.~H.,  2011, \mn@doi [\aj] {10.1088/0004-6256/142/4/110}, \href
  {http://adsabs.harvard.edu/abs/2011AJ....142..110M} {142, 110}

\bibitem[\protect\citeauthoryear{{Michel} et~al.,}{{Michel}
  et~al.}{2017}]{Michel2017}
{Michel} E.,  et~al., 2017, in EPJWC. p. 03001 (\mn@eprint {arXiv}
  {1705.03721})

\bibitem[\protect\citeauthoryear{{Murphy}, {Hey}, {Van Reeth}  \&
  {Bedding}}{{Murphy} et~al.}{2019}]{Murphy2019}
{Murphy} S.~J.,  {Hey} D.,  {Van Reeth} T.,   {Bedding} T.~R.,  2019, \mn@doi
  [\mnras] {10.1093/mnras/stz590}, \href
  {https://ui.adsabs.harvard.edu/abs/2019MNRAS.485.2380M} {485, 2380}

\bibitem[\protect\citeauthoryear{{Murphy}, {Bedding}, {Gautam}  \&
  {Joyce}}{{Murphy} et~al.}{2023}]{Murphy++2023}
{Murphy} S.~J.,  {Bedding} T.~R.,  {Gautam} A.,   {Joyce} M.,  2023, \mn@doi
  [\mnras] {10.1093/mnras/stad2849}, \href
  {https://ui.adsabs.harvard.edu/abs/2023MNRAS.526.3779M} {526, 3779}

\bibitem[\protect\citeauthoryear{{Palakkatharappil} \&
  {Creevey}}{{Palakkatharappil} \&
  {Creevey}}{2023}]{Palakkatharappil+Creevey2023}
{Palakkatharappil} D.~B.,  {Creevey} O.~L.,  2023, \mn@doi [\aap]
  {10.1051/0004-6361/202243624}, \href
  {https://ui.adsabs.harvard.edu/abs/2023A&A...674A.146P} {674, A146}

\bibitem[\protect\citeauthoryear{{Pamos Ortega}, {Garc{\'\i}a Hern{\'a}ndez},
  {Su{\'a}rez}, {Pascual Granado}, {Barcel{\'o} Forteza}  \&
  {Rod{\'o}n}}{{Pamos Ortega} et~al.}{2022}]{Pamos-Ortega++2022}
{Pamos Ortega} D.,  {Garc{\'\i}a Hern{\'a}ndez} A.,  {Su{\'a}rez} J.~C.,
  {Pascual Granado} J.,  {Barcel{\'o} Forteza} S.,   {Rod{\'o}n} J.~R.,  2022,
  \mn@doi [\mnras] {10.1093/mnras/stac864}, \href
  {https://ui.adsabs.harvard.edu/abs/2022MNRAS.513..374P} {513, 374}

\bibitem[\protect\citeauthoryear{{Pamos Ortega}, {Mirouh}, {Garc{\'\i}a
  Hern{\'a}ndez}, {Su{\'a}rez Yanes}  \& {Barcel{\'o} Forteza}}{{Pamos Ortega}
  et~al.}{2023}]{Pamos-Ortega++2023}
{Pamos Ortega} D.,  {Mirouh} G.~M.,  {Garc{\'\i}a Hern{\'a}ndez} A.,
  {Su{\'a}rez Yanes} J.~C.,   {Barcel{\'o} Forteza} S.,  2023, \mn@doi [\aap]
  {10.1051/0004-6361/202346323}, \href
  {https://ui.adsabs.harvard.edu/abs/2023A&A...675A.167P} {675, A167}

\bibitem[\protect\citeauthoryear{{Panda}, {Dhanpal}, {Murphy}, {Hanasoge}  \&
  {Bedding}}{{Panda} et~al.}{2024}]{Panda++2024}
{Panda} S.~K.,  {Dhanpal} S.,  {Murphy} S.~J.,  {Hanasoge} S.,   {Bedding}
  T.~R.,  2024, \mn@doi [\apj] {10.3847/1538-4357/ad0a97}, \href
  {https://ui.adsabs.harvard.edu/abs/2024ApJ...960...94P} {960, 94}

\bibitem[\protect\citeauthoryear{{Petersen} \&
  {Christensen-Dalsgaard}}{{Petersen} \&
  {Christensen-Dalsgaard}}{1996}]{Petersen+CD1996}
{Petersen} J.~O.,  {Christensen-Dalsgaard} J.,  1996, \aap, \href
  {https://ui.adsabs.harvard.edu/abs/1996A&A...312..463P} {312, 463}

\bibitem[\protect\citeauthoryear{{Poro} et~al.,}{{Poro}
  et~al.}{2021}]{Poro2021}
{Poro} A.,  et~al., 2021, \mn@doi [\pasp] {10.1088/1538-3873/ac12dc}, \href
  {https://ui.adsabs.harvard.edu/abs/2021PASP..133h4201P} {133, 084201}

\bibitem[\protect\citeauthoryear{{Riello} et~al.,}{{Riello}
  et~al.}{2021}]{Riello++2021}
{Riello} M.,  et~al., 2021, \mn@doi [\aap] {10.1051/0004-6361/202039587}, \href
  {https://ui.adsabs.harvard.edu/abs/2021A&A...649A...3R} {649, A3}

\bibitem[\protect\citeauthoryear{{Rix} \& {Bovy}}{{Rix} \&
  {Bovy}}{2013}]{Rix+Bovy2013}
{Rix} H.-W.,  {Bovy} J.,  2013, \mn@doi [\aapr] {10.1007/s00159-013-0061-8},
  \href {https://ui.adsabs.harvard.edu/abs/2013A&ARv..21...61R} {21, 61}

\bibitem[\protect\citeauthoryear{{Rodr{\'{\i}}guez}, {L{\'o}pez-Gonz{\'a}lez}
  \& {L{\'o}pez de Coca}}{{Rodr{\'{\i}}guez} et~al.}{2000}]{Rodriguez2000}
{Rodr{\'{\i}}guez} E.,  {L{\'o}pez-Gonz{\'a}lez} M.~J.,   {L{\'o}pez de Coca}
  P.,  2000, \aaps, \href {http://adsabs.harvard.edu/abs/2000A%26AS..144..469R}
  {144, 469}

\bibitem[\protect\citeauthoryear{{Rybizki} et~al.,}{{Rybizki}
  et~al.}{2022}]{Rybizki++2022}
{Rybizki} J.,  et~al., 2022, \mn@doi [\mnras] {10.1093/mnras/stab3588}, \href
  {https://ui.adsabs.harvard.edu/abs/2022MNRAS.510.2597R} {510, 2597}

\bibitem[\protect\citeauthoryear{{Scutt}, {Murphy}, {Nielsen}, {Davies},
  {Bedding}  \& {Lyttle}}{{Scutt} et~al.}{2023}]{Scutt++2023}
{Scutt} O.~J.,  {Murphy} S.~J.,  {Nielsen} M.~B.,  {Davies} G.~R.,  {Bedding}
  T.~R.,   {Lyttle} A.~J.,  2023, \mn@doi [\mnras] {10.1093/mnras/stad2621},
  \href {https://ui.adsabs.harvard.edu/abs/2023MNRAS.525.5235S} {525, 5235}

\bibitem[\protect\citeauthoryear{{Shi}, {Qian}  \& {Li}}{{Shi}
  et~al.}{2022}]{Shi++2022}
{Shi} X.-d.,  {Qian} S.-b.,   {Li} L.-J.,  2022, \mn@doi [\apjs]
  {10.3847/1538-4365/ac59b9}, \href
  {https://ui.adsabs.harvard.edu/abs/2022ApJS..259...50S} {259, 50}

\bibitem[\protect\citeauthoryear{{Skarka} et~al.,}{{Skarka}
  et~al.}{2022}]{Skarka++2022}
{Skarka} M.,  et~al., 2022, \mn@doi [\aap] {10.1051/0004-6361/202244037}, \href
  {https://ui.adsabs.harvard.edu/abs/2022A&A...666A.142S} {666, A142}

\bibitem[\protect\citeauthoryear{{Southworth}}{{Southworth}}{2021}]{Southworth2021}
{Southworth} J.,  2021, \mn@doi [Universe] {10.3390/universe7100369}, \href
  {https://ui.adsabs.harvard.edu/abs/2021Univ....7..369S} {7, 369}

\bibitem[\protect\citeauthoryear{{Steindl}, {Zwintz}  \&
  {M{\"u}llner}}{{Steindl} et~al.}{2022}]{Steindl++2022}
{Steindl} T.,  {Zwintz} K.,   {M{\"u}llner} M.,  2022, \mn@doi [\aap]
  {10.1051/0004-6361/202243242}, \href
  {https://ui.adsabs.harvard.edu/abs/2022A&A...664A..32S} {664, A32}

\bibitem[\protect\citeauthoryear{Terrell \& Scott}{Terrell \&
  Scott}{1992}]{Terrell+Scott1992}
Terrell G.~R.,  Scott D.~W.,  1992, \mn@doi [The Annals of Statistics]
  {10.1214/aos/1176348768}, 20, 1236

\bibitem[\protect\citeauthoryear{Virtanen et~al.,}{Virtanen
  et~al.}{2020}]{scipy2020}
Virtanen P.,  et~al., 2020, \mn@doi [Nature Methods]
  {10.1038/s41592-019-0686-2}, \href {https://rdcu.be/b08Wh} {17, 261}

\bibitem[\protect\citeauthoryear{{Xue}, {Niu}, {Xue}  \& {Yin}}{{Xue}
  et~al.}{2023}]{Xue++2023}
{Xue} W.,  {Niu} J.-S.,  {Xue} H.-F.,   {Yin} S.,  2023, \mn@doi [Research in
  Astronomy and Astrophysics] {10.1088/1674-4527/accdbc}, \href
  {https://ui.adsabs.harvard.edu/abs/2023RAA....23g5002X} {23, 075002}

\bibitem[\protect\citeauthoryear{{Ziaali}, {Bedding}, {Murphy}, {Van Reeth}  \&
  {Hey}}{{Ziaali} et~al.}{2019}]{Ziaali2019}
{Ziaali} E.,  {Bedding} T.~R.,  {Murphy} S.~J.,  {Van Reeth} T.,   {Hey} D.~R.,
   2019, \mn@doi [\mnras] {10.1093/mnras/stz1110}, \href
  {https://ui.adsabs.harvard.edu/abs/2019MNRAS.486.4348Z} {486, 4348}

\makeatother
\end{thebibliography}
\fi


\bsp	
\label{lastpage}
\end{document}